\documentclass[conference]{IEEEtran}
\IEEEoverridecommandlockouts
\usepackage{cite}
\usepackage{amsmath,amssymb,amsfonts}
\usepackage{algorithmic}
\usepackage{graphicx}
\usepackage{textcomp}
\def\BibTeX{{\rm B\kern-.05em{\sc i\kern-.025em b}\kern-.08em
    T\kern-.1667em\lower.7ex\hbox{E}\kern-.125emX}}

\usepackage{color}

\usepackage{multirow}
\usepackage[table,xcdraw]{xcolor}

\usepackage{paralist}
\usepackage{multirow}
\usepackage{hyperref}
\usepackage{balance}

\newcommand{\eg}{e.\,g.,\,}
\newcommand{\ie}{i.\,e.,\,}

\begin{document}

\title{Evaluating Deep Music Generation Methods\\ Using Data Augmentation
\thanks{This work is funded by the UK Economic \& Social Research Council (UK-ESRC) through the research Grant No.\ HJ-253479 (ACLEW), the Engineering and Physical Sciences Research Council (EPSRC) Grant No.\ 2021037, and the the DFG’s Reinhart Koselleckproject No. 442218748 (AUDI0NOMOUS). \\{978-1-6654-3288-7/21/\$31.00 ©2021 IEEE}}
}


\author{\IEEEauthorblockN{Toby Godwin}
\IEEEauthorblockA{\textit{GLAM} \\
\textit{Imperial College London}\\
London, UK \\
tobygodwin1@gmail.com}
\and
\IEEEauthorblockN{Georgios Rizos}
\IEEEauthorblockA{\textit{GLAM} \\
\textit{Imperial College London}\\
London, UK \\
georgios.rizos12@imperial.ac.uk}
\and
\IEEEauthorblockN{Alice Baird}
\IEEEauthorblockA{\textit{EIHW} \\
\textit{University of Augsburg}\\
Augsburg, Germany \\
alice.baird@uni-a.de}
\and
\IEEEauthorblockN{Najla D. Al Futaisi}
\IEEEauthorblockA{\textit{GLAM} \\
\textit{Imperial College London}\\
London, UK \\
n.al-futaisi18@imperial.ac.uk}
\and
\IEEEauthorblockN{Vincent Brisse}
\IEEEauthorblockA{\textit{GLAM} \\
\textit{Imperial College London}\\
London, UK \\
v.brisse@gmail.com}
\and
\IEEEauthorblockN{Bj{\"o}rn W. Schuller}
\IEEEauthorblockA{\textit{GLAM} \\
\textit{Imperial College London}\\
London, UK \\
bjoern.schuller@imperial.ac.uk}
}

\maketitle

\begin{abstract}
Despite advances in deep algorithmic music generation, evaluation of generated samples often relies on human evaluation, which is subjective and costly. We focus on designing a homogeneous, objective framework for evaluating samples of algorithmically generated music. Any engineered measures to evaluate generated music typically attempt to define the samples' musicality, but do not capture qualities of music such as theme or mood. We do not seek to assess the musical merit of generated music, but instead explore whether generated samples contain meaningful information pertaining to emotion or mood/theme. We achieve this by measuring the change in predictive performance of a music mood/theme classifier after augmenting its training data with generated samples. We analyse music samples generated by three models -- SampleRNN, Jukebox, and DDSP -- and employ a homogeneous framework across all methods to allow for objective comparison. This is the first attempt at augmenting a music genre classification dataset with conditionally generated music. We investigate the classification performance improvement using deep music generation and the ability of the generators to make emotional music by using an additional, emotion annotation of the dataset. Finally, we use a classifier trained on real data to evaluate the label validity of class-conditionally generated samples.
\end{abstract}


\section{Introduction}
\label{sec:intro}

Recent advances in generative algorithms have been able to produce \cite{jukebox,hierarchical_wavenet, sampleRNN} realistic sounding music in the waveform domain, with large scale models now able to generate full length songs, including comprehensible lyrics \cite{jukebox}. While these models have produced impressive results, their evaluation has tended to rely on human evaluation \cite{hierarchical_wavenet,sampleRNN,DDSP:,wavenet, wave2midi},
which is inherently subjective and extremely costly to scale. Alternatively, evaluation has been approached by monitoring quantities related to reconstruction quality \cite{jukebox}, or engineered measures for evaluating musical merit, which are currently inadequate due to the vastness and diversity of the space of all music \cite{hierarchical_wavenet}.

The enjoyment of music is deeply personal. So instead of performing a potentially subjective, and difficult to scale human perception survey, or attempting to engineer a mathematical measure for comprehensively quantifying whether generated music is enjoyable or interesting to listen to, we want to perform a \textit{functional evaluation of class-conditionally generated music samples}: \ie we want to assess whether generated samples contain the targeted class information, focusing on mood/theme and emotion classes. We present a framework, based on data augmentation, for the evaluation of samples of generated music, that offers a homogeneous and fair treatment across generation methods. Specifically, we quantitatively assess and compare the extent to which generated samples contain meaningful information pertaining to music of a given mood/theme. We report the change in predictive performance of a machine classifier as an indicator of whether there is meaningful information in the generated samples related to the particular classification task.  We examine the generated samples in an additional manner; by using a machine classifier trained on the real training set to classify the generated samples, in order to validate whether they carry features that are similar to the real training data. Finally, we perform the same experiments on an almost comprehensive subset of our dataset with a relabelling, which we introduce, pertaining to coarser arousal/valence emotion classes. We use this tofurther investigate the properties of the generated samples and gain insights on the augmentation experimental performance.

We analyse the generated samples from three deep music generation methods of different philosophies: the autoregressive SampleRNN \cite{sampleRNN}, Vector Quantised VAE (VQ-VAE) \cite{vqvae} based Jukebox \cite{jukebox}, and Differentiable Digital Signal Processing (DDSP) \cite{DDSP:}. While the former two have been shown to generate commercial music~\cite{jukebox,sampleRNN,dadabots}, the latter was proposed in the context of monophonic audio. Here we explore its ability to reconstruct \textit{polyphonic music}. While each method was presented with some subjective evaluation (often by the authors), there has yet been little effort to quantitatively and objectively evaluate generated samples. We propose to do this via data augmentation. Finally, this is also the first time such music generation methods have been considered for data augmentation of musical classification.






In the following section we briefly discuss common methods for evaluating generated music as well as the generative models used in this study. In Section 3 we discuss the data we used. Then in Section 4 we outline the details of the evaluation framework and discuss results. In Section 5 make a further analysis of the generated samples and conclude in Section 6.

\section{Background}



The evaluation of music is an inherently challenging task given the subjective nature of music itself~\cite{music_evaluation, gan_metrics, parada2017perception}. It is common to employ multiple human annotators to evaluate each sample \cite{parada2017perception} in order to elicit confident insights, something that hinders scalability. Recent work~\cite{yang2020evaluation} proposes simple, musically informed metrics to evaluate the musicality of generated music. However these metrics do not capture abstract qualities of music such as its emotion or mood. In other research, evaluation of deep music generation tends to focus on subjective surveys~\cite{hierarchical_wavenet, DDSP:}, or the reporting of reconstruction loss related quantities \cite{jukebox}. 


In the emotional speech synthesis domain (another subjective audio domain), the  generative adversarial network based studies performed in~\cite{bao2019cyclegan, rizos2020stargan, baird2019can} quantitatively evaluated synthesised speech using a classifier to demonstrate that the generated speech contained meaningful emotional information. We draw from these techniques to assess and compare the ability of three generative models from different deep learning paradigms to generate music that contains meaningful information relating to mood/theme or a listener's emotional response.


\subsection{Generative Models for Music}

There have been significant advancements in the field of music generation, particularly in the waveform domain~\cite{jukebox,hierarchical_wavenet,DDSP:, wavenet,wave2midi,pixelRNN,music_transformer}. We functionally evaluate the performance of three generative models, which subscribe to different generation paradigms.


\textbf{SampleRNN. }
The autoregressive SampleRNN~\cite{sampleRNN} uses tiers of RNN modules that work on different timescales of the signal. Higher level tiers process the signal at lower temporal resolutions to lower level tiers, and each tier is conditioned on a  vector from the tier above, which contains higher level contextual information.
The tiered architecture of the SampleRNN allows for different computational focus to be applied to different levels of abstraction of the audio, which allows long term dependencies to be modelled efficiently. 

\textbf{Jukebox. }
Jukebox \cite{jukebox} trains three separate convolutional VQ-VAEs \cite{vqvae} at different levels of abstraction.
Each level learns a discrete latent codebook of the input, so Jukebox learns three separate representations of the input data. Prior distributions of the latent codes are approximated using autoregresive methods \cite{sparse_transformer}, and music is generated by sampling from the latent codebooks, upsampling the lower resolution codes and decoding a single high-resolution latent representation to produce music.

\textbf{DDSP. } DDSP \cite{DDSP:} learns the parameters of deterministic digital signal processing techniques; used then to synthesise audio. The model itself is an autoencoder acting on Mel-spectrograms, where the latent representation is decomposed into: the time-varying fundamental frequency of the audio $F_0$, loudness $l$, and a latent vector that encodes the input $z$. All latent representations are time-varying and sequential, and  since $F_0$ and $l$ have interpretable, physical meanings. Here, we only consider the harmonic plus noise synthesiser model \cite{synthesiser,synthesiser2}.
 
\section{MTG-Jamendo Dataset}

For our experiments we use the MTG-Jamendo dataset \cite{mtg-jamendo-dataset}, which is a large collection of labelled, high-quality commercial music. We use a subset of MTG-Jamendo, labelled according to mood/theme, which was used in the MediaEval 2019 music classification competition \cite{bogdanov2019mediaeval}. There are a total of 56 well-balanced classes ranging from `epic' to `dance'. In order to gain insights respective to which types of music the generation models can generate, we utilise a second label set: this is an almost fully overlapping subset of the first, on which we performed a more coarse relabelling based on a psychology-based interpretation of the original labels. A summary of the duration of data partitions is given in Table \ref{tab:dataset} for both sets of labels.  The reader is referred to~\cite{mtg-jamendo-dataset, bogdanov2019mediaeval} for statistics of data with mood/theme labels.

\begin{table}[t]
\centering
\begin{tabular}{l|r|r|r|r}
                    & \textbf{Train} & \textbf{Val} & \textbf{Test} & $\sum$ \\ \hline
\textbf{Mood/theme} & 160            & 22           & 22            & 204                          \\
\textbf{Emotional}  & 157            & 20           & 20            & 197                         
\end{tabular}
%
\caption{Train, Val(idation), and Test partition durations (hours) for mood/theme and emotional labels. }
\label{tab:dataset}

\end{table}

\subsection{Music Emotion Labels}


We derive the emotional labels from a theory of emotion that takes into account a person's categorical emotion responses in terms of arousal and valence \cite{emotional_valence}. Valence is a psychological term that describes the intrinsic pleasure that a person derives from something and arousal describes the amount of attention that a person pays to something. Therefore, the labels (`activated pleasant', `activated unpleasant', `deactivated pleasant', `deactivated unpleasant') describe the expected arousal and valence responses of an individual when listening to the music. We label the music by grouping data such that the mood/theme labels are mapped to quadrants of the circumplex model of emotions \cite{emotional_valence} (such coarse binning of arousal and valence into classes is not uncommon \cite{zhang2019attention}): \eg since `happiness' is mapped to activated pleasant according to the model, tracks with the `happy' mood/theme label are labelled as `activated pleasant'. This process is extended for all mood/theme labels that are clearly mapped to evoked arousal and valence and we include the mapping thereof in the project webpage\footnote{\url{https://github.com/glam-imperial/Functional-Music-Generation-Evaluation}}.

\begin{figure}[t]
%
\begin{minipage}[b]{0.9\linewidth}
  \centering
  \centerline{\includegraphics[width=0.75\linewidth]{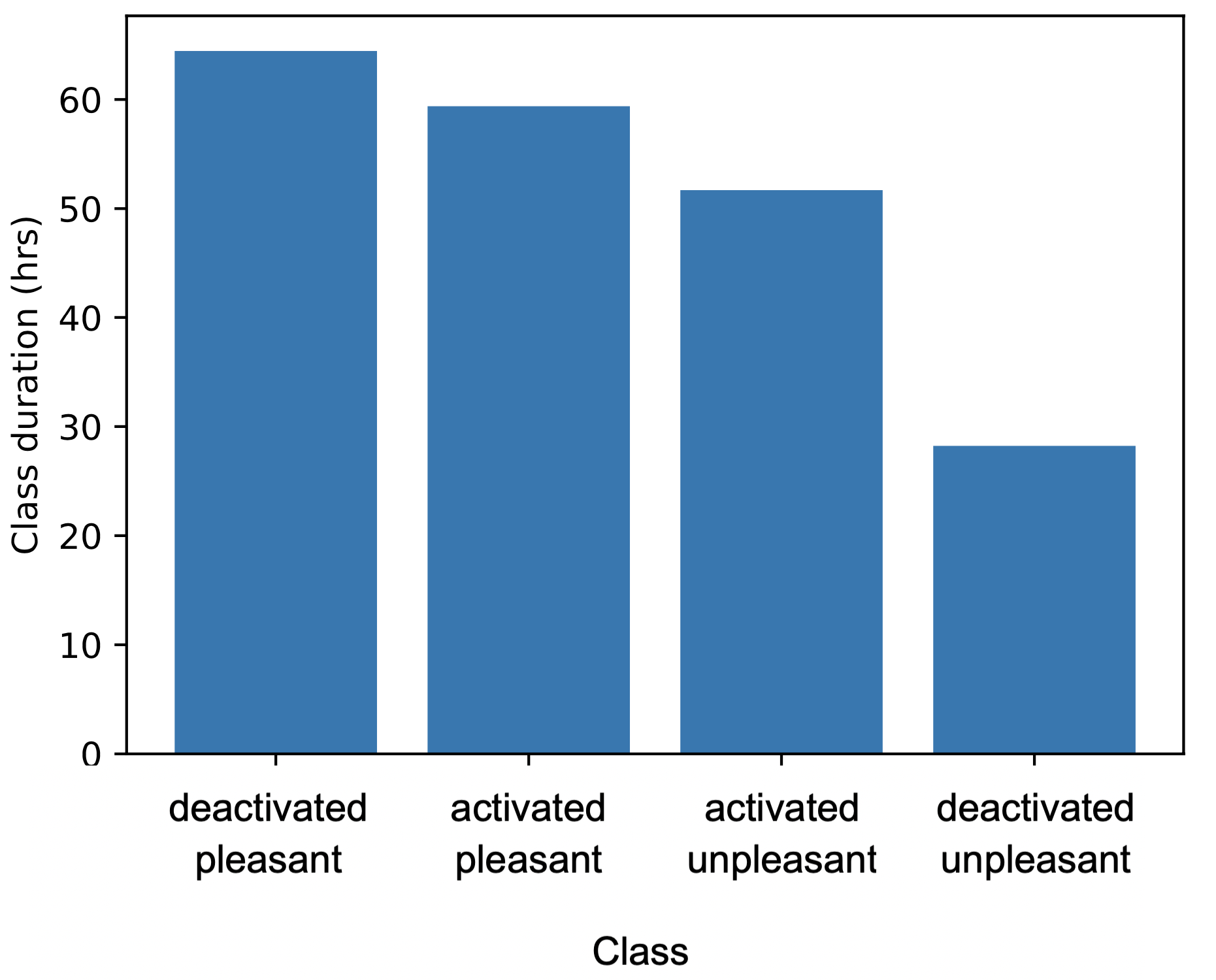}}

\end{minipage}
\caption{Distribution of total duration of music for each emotional class for the dataset used in this study. Whereas there is a degree of class imbalance we note that it is not overwhelming. 
}
\label{fig:classes_distribution}
\end{figure}


\section{Evaluation Framework}
\begin{figure*}[htbp]
\begin{minipage}[b]{.32\linewidth}
  \centering
  \centerline{\includegraphics[width=5.93cm]{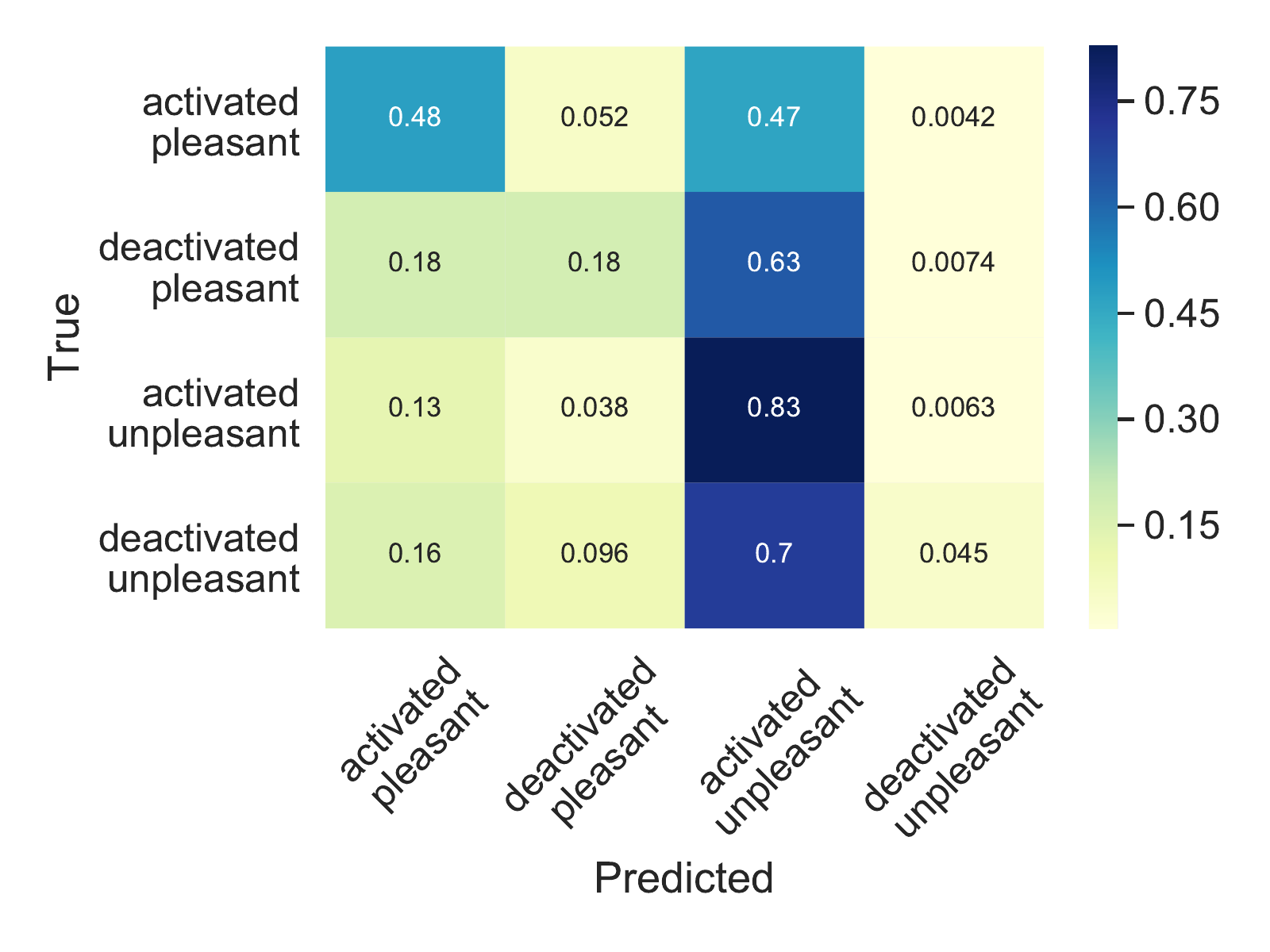}}
  \centerline{\small{(a) SampleRNN}}
\end{minipage}
\begin{minipage}[b]{.32\linewidth}
  \centering
  \centerline{\includegraphics[width=5.93cm]{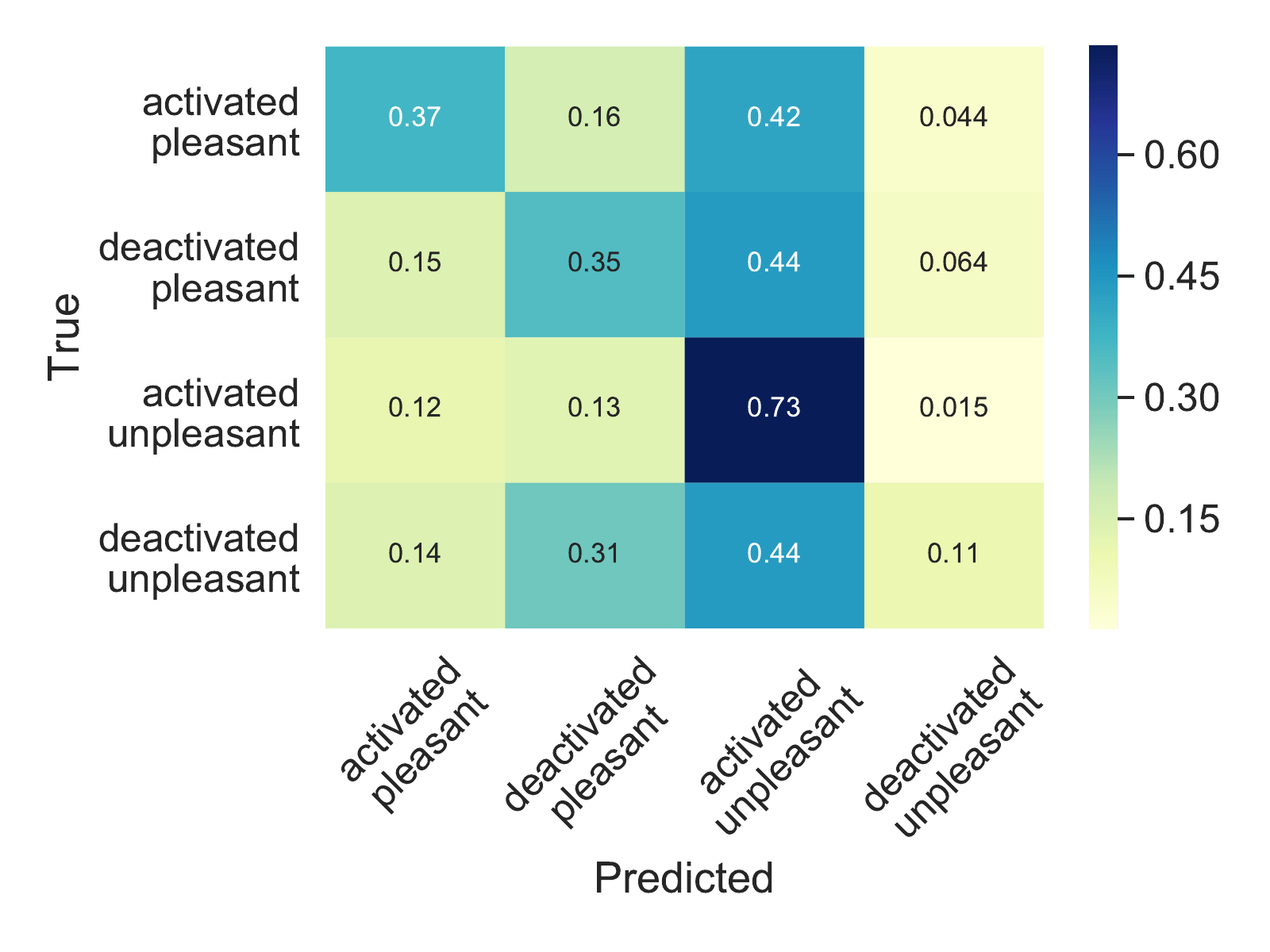}}
  \centerline{\small{(b) Jukebox}}
\end{minipage}
\begin{minipage}[b]{0.32\linewidth}
  \centering
  \centerline{\includegraphics[width=5.93cm]{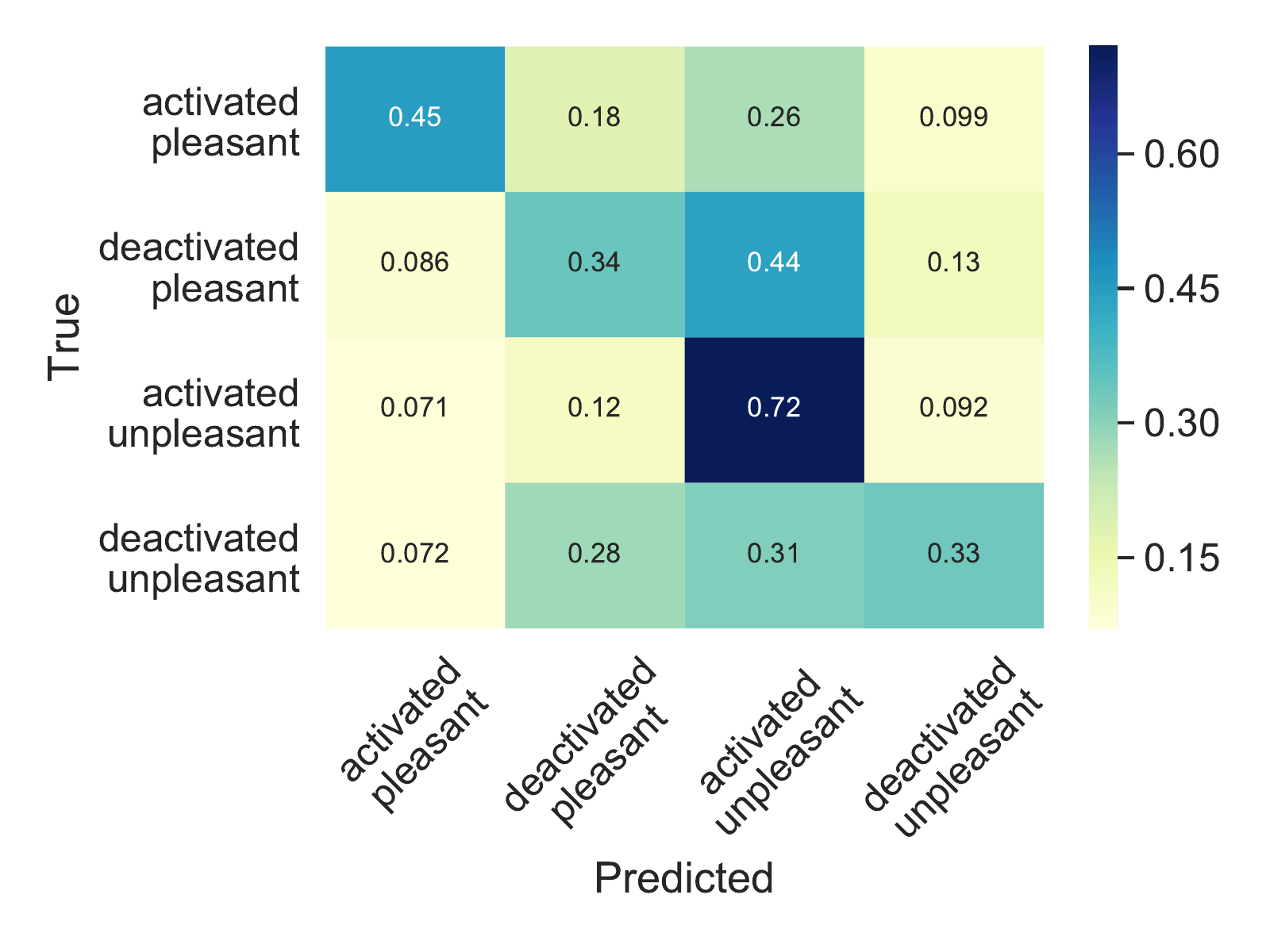}}
  \centerline{\small{(c) DDSP}}
\end{minipage}
\caption{Normalised confusion matrices for classification of generated samples with emotional labels. The three music generation methods tend to produce samples that are perceived by the classifier as `activated-unpleasant'. DDSP has a stronger diagonal in the confusion matrix, and is the only one that achieves good performance in predicting the `deactivated-unpleasant' class.}
\label{fig:class_emotion}
\end{figure*}
We assess whether the generated samples contain meaningful information pertaining to music of a given mood/theme or emotion by measuring whether their usage for data augmentation increases the predictive performance of a classifier on two tasks (mood/theme and emotional response). We compare classification performance after augmentation to the performance of a baseline classifier trained only on the training partition. The relatively strong baseline classification performance implies that the abstract features of thematic and emotional music are indeed modelled by the classifier. Therefore, we can say that an improvement over the baseline after augmentation indicates that, indeed, the samples contain meaningful information relating to music of a given mood/theme or emotion. 

We measure classifier performance according to the three metrics that were used in the MediaEval 2019 competition: F1-score, area under the precision-recall curve (PR-AUC), and the area under the Receiver Operator Characteristics curve (ROC-AUC). We adopt two means of averaging each metric: a) micro-averaged, which is the harmonic mean of the overall metric scores and b) macro-averaged, which performs an unweighted average of class-specific metric scores.  We evaluate performance on the test partition of each dataset with respect to micro and macro averaged metrics, calculated in the same manner as the MediaEval 2019 competition submission.


We use a classifier architecture \cite{classifier_paper} that came fourth in the MediaEval 2019 competition with respect to macro-averaged PR-AUC. The model was developed specifically for the data used in this work, and uses a pretrained MobileNetV2 \cite{mobilenetv2} block in combination with a self attention block to classify music from its Mel-spectrogram representation. Since this classifier was developed specifically for the challenge, we used the same hyperparameters reported in its submission paper \cite{classifier_paper}.

\subsection{Augmentation Policy}

\begin{table}[t]
\small
\centering
\resizebox{\columnwidth}{!}{%
\begin{tabular}{l|r|r|r|r}
\multicolumn{1}{r|}{} & \textbf{\begin{tabular}[c]{@{}c@{}}Sample\\ type\end{tabular}} & \textbf{\begin{tabular}[c]{@{}c@{}}Sample\\ length (s)\end{tabular}} & \textbf{\begin{tabular}[c]{@{}c@{}}Prime\\  length (s)\end{tabular}} & \textbf{\begin{tabular}[c]{@{}c@{}}Sample\\  rate (kHz)\end{tabular}} \\ \hline
\textbf{SampleRNN}    & Primed                                                         & 10                                                                   & 4                                                                    & 16                                                                    \\
\textbf{DDSP}         & Recons                                                         & 4                                                                    & n/a                                                                  & 16                                                                    \\
\textbf{Jukebox}      & Primed                                                         & 24                                                                   & 8                                                                    & 44                                                                   
\end{tabular}
}
\caption{Model properties pertaining to generating samples.}
\label{tab:sample_properties}
\end{table}

For each generative method, we generate an equal duration of music per class, with the total length of generated music equal to 5\,\% of the duration of the train split of each dataset (8 hours for mood/theme, 7.85 hours for emotional). We then augment each class with the fixed duration of generated music. This gives a balanced treatment per class, increases the likelihood of successfully observing the augmentation effect, and allows for a fair comparison across generative methods. 
Since each method generates samples of different lengths (Table \ref{tab:sample_properties}), we adjust the number of samples accordingly to generate a fixed duration of music per class.
We trained SampleRNN and DDSP on the training partition of each dataset and used performance on the validation partition to tune hyperparameters. We did not retrain Jukebox, and opted instead to use the pre-trained model from \cite{jukebox}, trained on 1.2 million songs collected from the web.

We generate samples with SampleRNN and Jukebox by `priming' the model with an input length of real music to seed the sampling process. We select a prime length for each method based on its desired input and maximum capable sample length.  Generated music comprises the majority of each sample, however, the priming sample length is not insignificant, which we believe justifies our assumption that the completed sequence may inherit the label from the input music for data augmentation. `Priming' samples are used once, randomly sampled per class from the training partition without replacement.

As for DDSP, we reconstruct the input music sample and assume that the reconstructed sample has the same label as the original. The input music is again randomly sampled from the training partition without replacement. Although DDSP was designed for monophonic music, it can also reconstruct polyphonic music since it is trained to reconstruct a spectrogram. 


\section{Data Augmentation Experiments}

Table~\ref{tab:augmentation_results_default} summarises our data augmentation experimental results on both label types. All values reported are averaged across two trials. We observe from Table \ref{tab:augmentation_results_default}(a) that DDSP's samples yielded a consistent increase in performance with respect to micro and macro averaged metrics; a quantitative indication that these samples contain meaningful information.  Comparing our performance to the blind evaluation of the MediaEval 2019 competition would yield an absolute improvement to the baseline submission of 
0.8\,\% with respect to macro-averaged PR-AUC and ROC-AUC, promoting the method from 4th to 3rd place with respect to macro averaged ROC-AUC and PR-AUC. This is an important improvement, especially given that the performance achieved by our replication of the classifier proposed in \cite{classifier_paper} was lower than the authors' reported \textit{submission} performance, particularly with respect to macro-averaged ROC-AUC. 
Samples generated by SampleRNN and Jukebox failed to result in consistent performance increases to the same degree, although they were at least as good as the baseline in terms of PR-AUC and ROC-AUC, and competitive in terms of F1. One hypothesis for this behaviour is that these are both priming based generation methods: the final generated sample may deviate significantly from the priming sample class, impacting the augmentation behaviour in this numerous class setting. 

\begin{table}[!t]

\centering
\resizebox{\columnwidth}{!}{%
\begin{tabular}{|c|c|c|c|c|c|c|}
\hline
                                 & \multicolumn{2}{c|}{\textbf{F1}}                              & \multicolumn{2}{c|}{\textbf{PR-AUC}}                          & \multicolumn{2}{c|}{\textbf{ROC-AUC}}                         \\ \cline{2-7} 
\multirow{-2}{*}{\textbf{Model}} & \textbf{Macro}                & \textbf{Micro}                & \textbf{Macro}                & \textbf{Micro}                & \textbf{Macro}                & \textbf{Micro}                \\ \hline
SampleRNN                        & 0.063                         & \cellcolor[HTML]{B7E1CD}0.137 & \cellcolor[HTML]{B7E1CD}0.134 & \cellcolor[HTML]{B7E1CD}0.158 & 0.752                         & 0.799                         \\ \hline
DDSP                             & \cellcolor[HTML]{B7E1CD}0.066 & \cellcolor[HTML]{B7E1CD}0.146 & \cellcolor[HTML]{B7E1CD}0.134 & \cellcolor[HTML]{B7E1CD}0.159 & \cellcolor[HTML]{B7E1CD}0.761 & \cellcolor[HTML]{B7E1CD}0.807 \\ \hline
Jukebox                          & 0.063                         & 0.133                         & 0.125                         & \cellcolor[HTML]{B7E1CD}0.151 & 0.744                         & 0.795                         \\ \hline

\end{tabular}
}
\centerline{(a) Mood/theme labels }
\centerline{\tiny{}  }
\resizebox{\columnwidth}{!}{%
\begin{tabular}{|c|c|c|c|c|c|c|}
\hline
                                 & \multicolumn{2}{c|}{\textbf{F1}}                              & \multicolumn{2}{c|}{\textbf{PR-AUC}}                          & \multicolumn{2}{c|}{\textbf{ROC-AUC}}                         \\ \cline{2-7} 
\multirow{-2}{*}{\textbf{Model}} & \textbf{Macro}                & \textbf{Micro}                & \textbf{Macro}                & \textbf{Micro}                & \textbf{Macro}                & \textbf{Micro}                \\ \hline
SampleRNN                        & \cellcolor[HTML]{FFFFFF}0.483 & \cellcolor[HTML]{FFFFFF}0.523 & \cellcolor[HTML]{FFFFFF}0.515 & \cellcolor[HTML]{FFFFFF}0.565 & \cellcolor[HTML]{FFFFFF}0.764 & \cellcolor[HTML]{FFFFFF}0.774 \\ \hline
DDSP                             & \cellcolor[HTML]{FFFFFF}0.489 & \cellcolor[HTML]{FFFFFF}0.526 & \cellcolor[HTML]{FFFFFF}0.511 & \cellcolor[HTML]{FFFFFF}0.565 & \cellcolor[HTML]{FFFFFF}0.759 & \cellcolor[HTML]{FFFFFF}0.771 \\ \hline
Jukebox                          & \cellcolor[HTML]{FFFFFF}0.486 & \cellcolor[HTML]{B7E1CD}0.532 & \cellcolor[HTML]{FFFFFF}0.524 & \cellcolor[HTML]{B7E1CD}0.587 & \cellcolor[HTML]{FFFFFF}0.771 & \cellcolor[HTML]{B7E1CD}0.783 \\ \hline
\end{tabular}
}
\centerline{(b) Emotional (arousal/valence) labels}
\centerline{\tiny{} }
\resizebox{\columnwidth}{!}{%
\begin{tabular}{|c|c|c|c|c|c|c|}
\hline
\multirow{2}{*}{\textbf{Labels}} & \multicolumn{2}{c|}{\textbf{F1}} & \multicolumn{2}{c|}{\textbf{PR-AUC}} & \multicolumn{2}{c|}{\textbf{ROC-AUC}} \\ \cline{2-7} 
                                 & \textbf{Macro}  & \textbf{Micro} & \textbf{Macro}    & \textbf{Micro}   & \textbf{Macro}    & \textbf{Micro}    \\ \hline
Mood/theme (ours)                       & 0.064           & 0.134          & 0.125             & 0.148            & 0.745             & 0.794             \\ \hline
Emotional                        & 0.485           & 0.523          & 0.524             & 0.574            & 0.764             & 0.771             \\ \hline
\end{tabular}
}
\centerline{(c) Baseline classification performance.}
\caption{Predictive performance of baseline classifier on test partition, and test partition classification results after  data augmentation. We highlight the cells that have at least 1\,\% increase in performance relative to the performance of our replication of the baseline classifier.}
\label{tab:augmentation_results_default}
\end{table}


\begin{figure*}[!t]
\begin{minipage}[b]{.32\linewidth}
  \centering
  \centerline{\includegraphics[width=5.93cm]{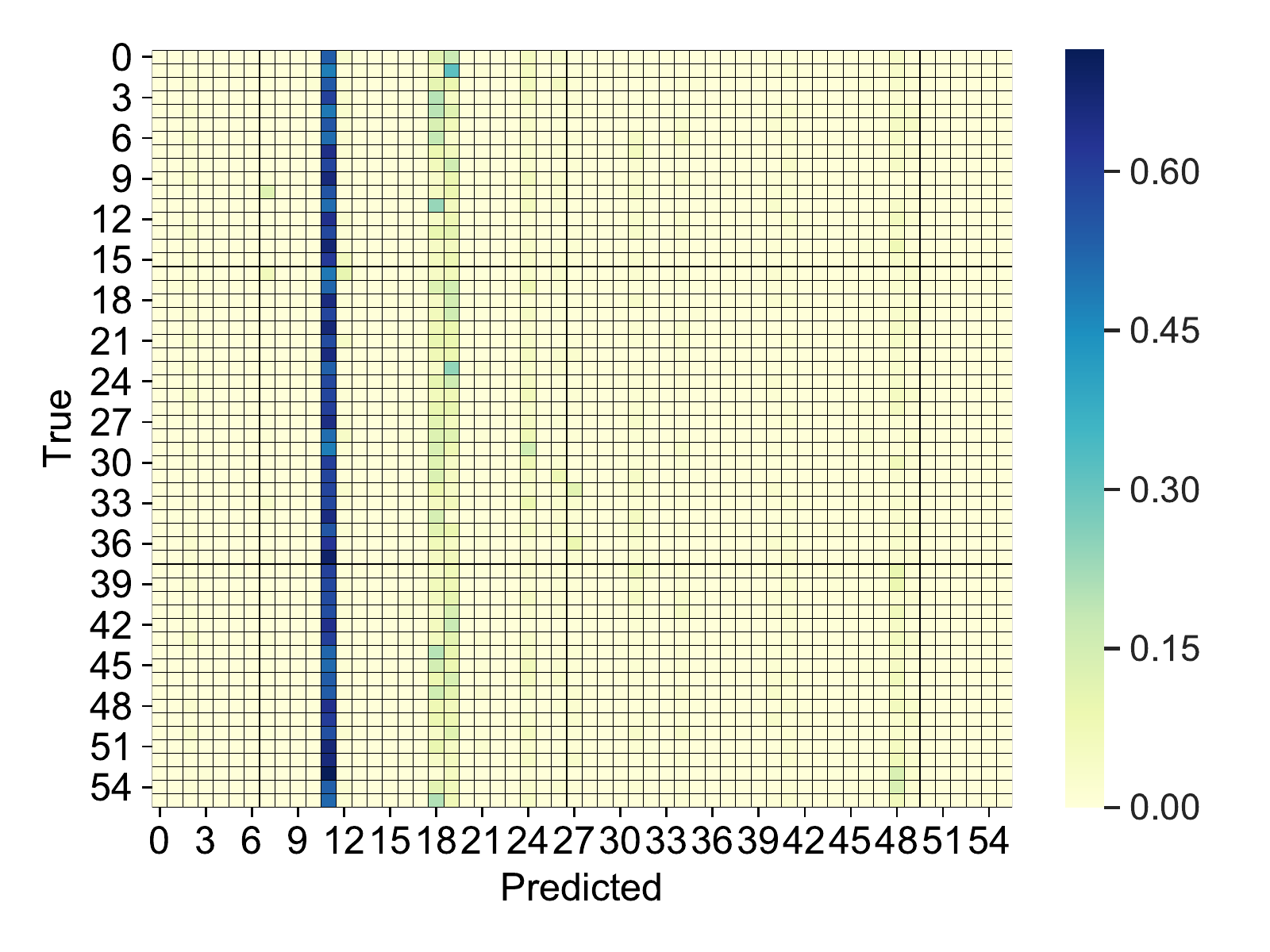}}
  \centerline{\small{(a) SampleRNN}}
\end{minipage}
\begin{minipage}[b]{.32\linewidth}
  \centering
  \centerline{\includegraphics[width=5.93cm]{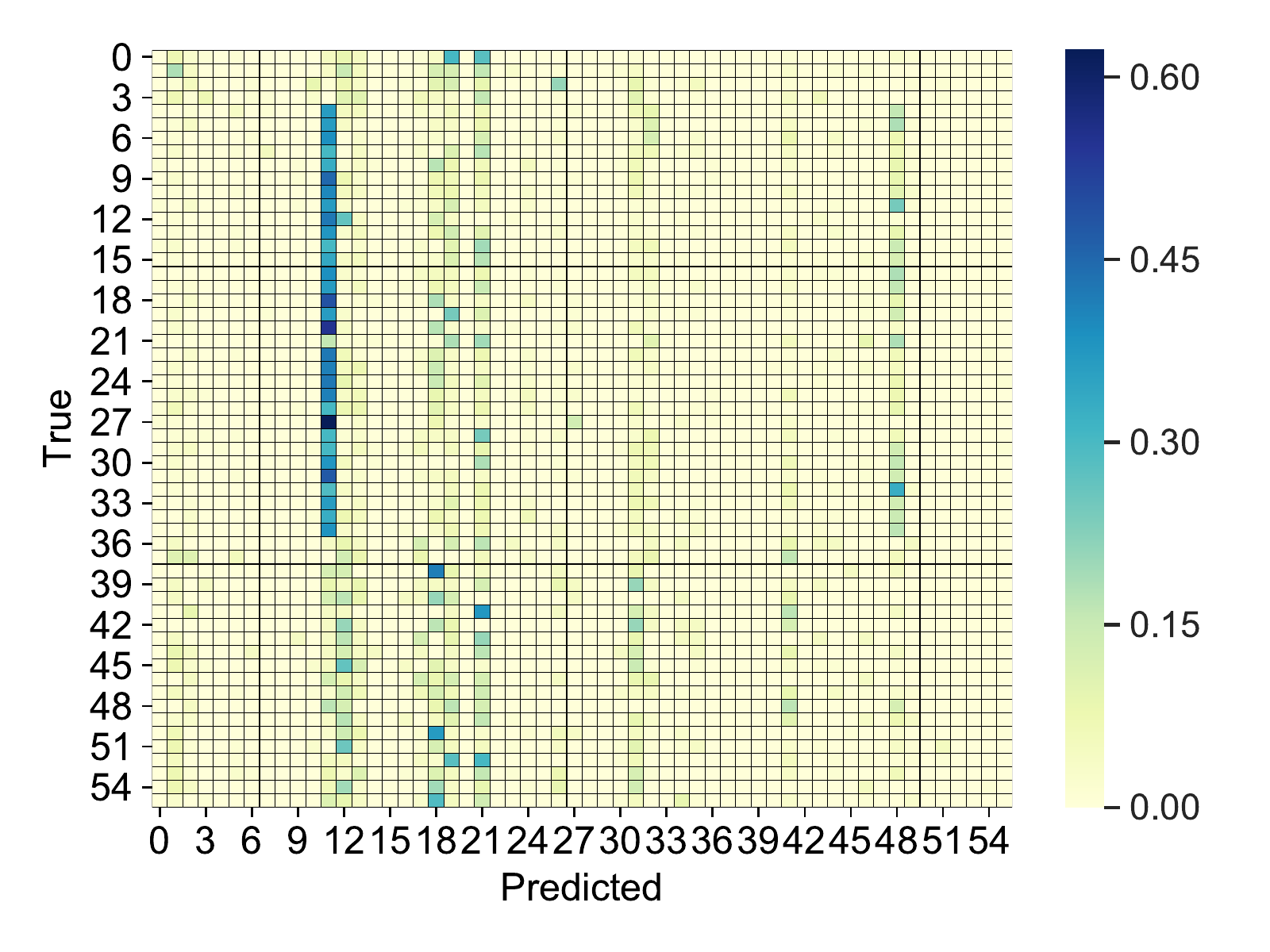}}
  \centerline{\small{(b) Jukebox}}
\end{minipage}
\begin{minipage}[b]{.32\linewidth}
  \centering
  \centerline{\includegraphics[width=5.93cm]{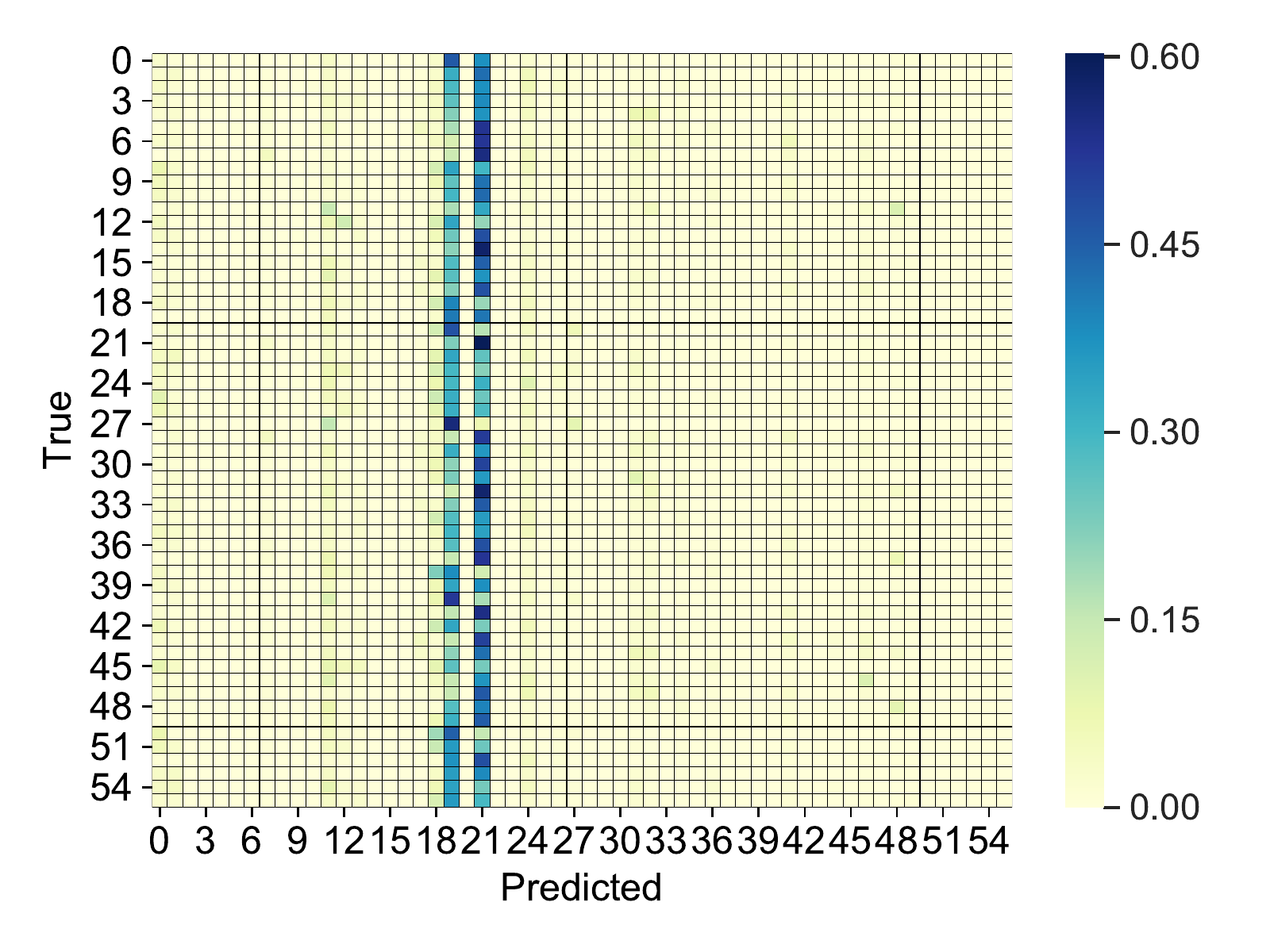}}
  \centerline{\small{(c) DDSP}}
\end{minipage}
\caption{Normalised confusion matrices for classification of generated samples with mood/theme labels. The SampleRNN samples tend to be predicted as belonging to the `dark' class. Greater range of possible predicted classes is observed for the Jukebox samples, albeit `dark' is still somewhat dominant. The DDSP samples are predicted mostly as `epic' or `film'. We see that existing music generation methods implicitly bias the generated samples with specific thematic indices.}
\label{fig:class_default}
\end{figure*}

For emotional labels, we observe from Table \ref{tab:augmentation_results_default}(b) that samples generated by SampleRNN and DDSP yield negligible performance increase. Only Jukebox samples yield consistent performance increase with respect to micro-averaged metrics, implying that performance was unequal across classes. We investigated and found that across all trials, only the `activated pleasant' class consistently improved the classifier after augmentation. This implies that all methods could only generate music with meaningful content from `activated pleasant' music. 


\section{Classification of Generated Samples}

We use the pre-trained baseline classifier to identify whether there is consistency between dominant features in the generated samples and characteristic features in the real training data. We measure the testing predictive performance of the classifier on the generated samples.
We observe from Figure \ref{fig:class_emotion} that samples tend to be classified as  `activated unpleasant'. All samples contain a degree of noise, and we hypothesise that this noise correlates with characteristic features of `activated unpleasant' music. Similarly for mood/theme labels, Figure \ref{fig:class_default} shows that all samples from all methods tend to be classified as having a `dark' mood or theme (`11' in Figure \ref{fig:class_default}).
 The classifier predicts a greater variety of classes for both DDSP and Jukebox, implying that samples generated by these methods contain features that correlate to characteristic features of a range of classes.  Figure \ref{fig:class_default} shows that DDSP's mood/theme samples tend to be classified as `epic' or `film'. Since DDSP reconstructs music from deterministic 
 estimates of loudness, elements such as  percussion, and changes in dynamics are often well captured in reconstructed samples, which we believe are characteristic features of `epic' and `film' music.

\section{Conclusion \& Future Work}
Figure~\ref{fig:class_default} reveal that no model could generate music with meaningful features for \textit{all} classes of each dataset. We further show that each model is particularly effective at generating meaningful music for certain classes, indicating that each model has a `character' in terms of the music it generates. Figure~\ref{fig:class_emotion} indicates that DDSP generalised best to all classes, implying that polyphonic music, reconstructed by DDSP, maintains much of its meaningful information. We demonstrate, by comparison to the MediaEval 2019 competition, that music genre classifiers are improved by using reconstructed samples from DDSP (less so from SampleRNN) to augment training data; however, this effect was not observed in the emotion classification experiment, where only JukeBox achieved some improvement in Micro averages of the performance measures used in MediaEval 2019.

We firmly believe that objective evaluation of music generation methods is key, in order to fairly monitor progress in this domain.
Our experiments show that all models struggled to generate music with dominant features that were similar to characteristic features of a given theme/mood class, possibly due to artefacts present in the generated music.

This analysis depends on the behaviour of the classifier, so future work should explore the effect of different classifier architectures. Since the generated samples will inevitably be from a different distribution to the training data, using domain adversarial training \cite{domain_adversarial, domain_emotion_recog} would help to uncover domain invariant features in the generated samples, and is therefore an interesting avenue for further research. We have also assumed that generated samples strictly inherit the label from an original or priming sample: as we have seen, this is not necessarily the case, with only the DDSP method being consistent in its class-conditional generation, for the emotion prediction task. To that end, we believe that a more elaborate consideration of the confidence we have regarding the augmentation labels is required, and intend to explore this matter with regularisation methods like virtual adversarial training \cite{miyato2018virtual}.

\balance

\bibliographystyle{IEEEtran}
\bibliography{IEEEabrv,mybib}

\end{document}